\documentclass[%
superscriptaddress,
nofootinbib,
 amsmath,
amssymb,
prl,
reprint,
]{revtex4-2}

\usepackage[T1]{fontenc}
\usepackage{graphicx}
\usepackage{dcolumn}
\usepackage{bm}
\usepackage{braket}
\usepackage{mathtools}
\usepackage{hyperref}
\usepackage{xcolor}

\newcommand{\correct}{\text{c}}
\newcommand{\incorrect}{\text{i}}
\newcommand{\discard}{\text{d}}

\begin{document}
\title{Efficient Concatenated Bosonic Code for Additive Gaussian Noise}

\author{Kosuke Fukui} 
\affiliation{Department of Applied Physics, School of Engineering, The University of Tokyo,\\
7-3-1 Hongo, Bunkyo-ku, Tokyo 113-8656, Japan}
\author{Takaya Matsuura}
\affiliation{Centre for Quantum Computation and Communication Technology,
School of Science, RMIT University, Melbourne, VIC 3000, Australia}
\author{Nicolas C. Menicucci}
\affiliation{Centre for Quantum Computation and Communication Technology,
School of Science, RMIT University, Melbourne, Victoria 3000, Australia}

\begin{abstract}
Bosonic codes offer noise resilience for quantum information processing.
Good performance often comes at a price of complex decoding schemes, limiting their practicality.
Here, we propose using a Gottesman-Kitaev-Preskill code to detect and discard error-prone qubits, concatenated with a quantum parity code to handle the residual errors. Our method employs a simple, linear-time decoder that nevertheless 
offers significant performance improvements over the standard decoder. 
Our work may have applications in a wide range of quantum computation and communication scenarios.
\end{abstract}

\maketitle
{\it Introduction.}---%
Quantum-error-correcting codes known as {bosonic codes}~\cite{Albert2018} protect discrete quantum information encoded in bosonic mode(s). The infinite-dimensional nature of the bosonic Hilbert space allows more sophisticated encoding than the conventional single-photon encoding~\cite{Knill2001,Kok2007,Rudolph2017} or matter-based qubits~\cite{Kjaergaard2020,Koch2007,Schreier2008}. Various codes exist that protect encoded quantum information against decoherence, with better than break-even performance demonstrated recently~\cite{Ofek2016,Hu2019}. 
These versatile codes find applications in optical, solid-state, and vibrational systems. We call qubits encoded in a bosonic code {bosonic qubits}.

While quantum supremacy has been demonstrated in both solid-state qubits~\cite{Arute2019} and optics~\cite{Zhong2020,madsen2022quantum}, the ultimate goal of a large-scale, fault-tolerant quantum computer will require additional innovations, and its ultimate architecture remains an open question. Such requirements on the device can be roughly classified into scalability (many qubits) and fault tolerance (of good quality)~\cite{DiVincenzo2000}, and architectures designed to use bosonic qubits as the information carriers have recently demonstrated prominent advances in both areas.

Progress on scalability has been most significant in optics through demonstrations of computationally universal continuous-variable~(CV) cluster states~\cite{Menicucci2006,Menicucci2011a} comprising ${\sim 10^4}$ modes~\cite{Asavanant2019,Larsen2019a} and measurement-based implementation of CV quantum gates~\cite{Asavanant2020,Larsen2020}. When used to process the bosonic qubits proposed by Gottesman, Kitaev, and Preskill~(GKP)~\cite{Gottesman2001}---and with high enough squeezing---these architectures can be made fault tolerant~\cite{Menicucci2014, Walshe2019}.

The GKP qubit~\cite{Gottesman2001} has emerged as a promising bosonic qubit for fault tolerance due to its excellent performance against common types of noise~\cite{Albert2018}. Experiments involving trapped ions~\cite{Fluhmann2019,deNeeve2022} and superconducting circuits~\cite{Campagne-Ibarcq2020a,Sivak2023} have demonstrated a GKP qubit, with the latter boasting a squeezing level close to 10~dB. This level is sufficient for fault tolerance in some proposed architectures~\cite{Fukui2018,Fukui2019} and is approaching what is required by others~\cite{Bourassa2020,tzitrin2021fault,Larsen2021}. Numerous proposals exist to produce these states in optics as well~\cite{pirandola2004constructing, pirandola2006continuous, pirandola2006generating, motes2017encoding,eaton2019non,su2019conversion,
arrazola2019machine,tzitrin2020progress,lin2020encoding,hastrup2022protocol,fukui2022generating,fukui2022efficient,takase2022gaussian,yanagimoto2023quantum}.
Furthermore, the GKP qubit performs well for quantum communication~\cite{Bennett1984,Briegel1998,Kimble2008} thanks to the robustness against photon loss~\cite{Albert2018}. In fact, recent results show that using GKP qubits may greatly enhance long-distance quantum communication~\cite{Fukui2020,Rozpkedek2020}.

A conventional noise model in bosonic systems is the Gaussian quantum channel~(GQC)~\cite{Harrington2001,Gottesman2001}, also known as additive Gaussian noise. While the landscape of Gaussian operations includes other types of channels~\cite{Caruso2006,Holevo2007}, the GQC is a particularly common one~\cite{Eisert2007}, and we focus on it here.
The GQC is a simple, canonical type of noise for analyzing bosonic code performance~\cite{Gottesman2001,Harrington2001}. Buoyed by the fact that displacements form an operator basis, protecting against the GQC allows some level of protection against all types of bosonic noise~\cite{Gottesman2001}. 
The GKP encoding is specifically designed to protect qubit information against the GQC (and thus against bosonic noise in general), but despite this, its performance ``out of the box'' as a single-mode code is suboptimal against the GQC~\cite{Harrington2001,Gottesman2001}.

A long-standing open problem in CV quantum information~\cite{Harrington2001} is to design a {simple and efficient concatenated code} that enables finite-rate quantum communication with levels of GQC noise that are 
guaranteed to allow this~\cite{Holevo2001,Harrington2001}. This bound can be achieved by GKP-type codes based on high-dimensional sphere packing~\cite{Harrington2001}, but the authors of that work were unsatisfied with this because such a code is not concatenated and thus offers no obvious structure that might be exploited~\cite{Shor1996,DiVincenzo1998} to further improve its performance.

Analog quantum error correction~(QEC)~\cite{Fukui2017} makes strides toward achieving the goal of Ref.~\cite{Harrington2001} by using the real-valued syndrome of a GKP qubit to improve error recovery in a concatenated code by selecting the most likely error pattern for a given syndrome.
In fact, when used with a suitable qubit code, analog QEC can achieve the bound~\cite{Fukui2017,Fukui2018}. This would seem to be the end of the story except for one major drawback: 
The decoder for analog QEC employs a type of belief propagation~\cite{Poulin2006} that may become unwieldy in real-world implementations. This is especially true in optical architectures, where fast processing of the outcomes is vital~\cite{Bourassa2020,tzitrin2021fault}. In such cases---and especially when hardware-level control is used---reducing the number of bits required to represent outcomes may be critical to fast decoding. 

What we would like instead is a simple CV-level decoder that generates discrete outcomes that can be fed directly into a qubit-level code at the next level of concatenation. This is the key innovation that makes further improvements feasible since more complicated codes or additional layers of concatenation do not require modifying the CV-level decoding scheme, thus keeping the decoder simple and efficient.

In this Letter, we make significant progress toward achieving this goal.
Our innovation uses the CV-level measurement outcome from GKP error correction merely to decide whether to keep the qubit or discard it entirely and treat it as a located erasure error. This is a simple, local decoding step and does not require complicated modeling of CV-level errors. The quantum parity code (QPC)~\cite{Ralph2005} is well suited to dealing with the discarded qubits~\cite{Muralidharan2014,Munro2012,Ewert2016}, and we numerically show that concatenating the GKP code with a QPC considerably improves its performances with a small code and straightforward decoding,
linear in the number of modes.

{\it GKP qubit.}---%
The GKP code encodes a qubit in an oscillator in a way that protects against errors caused by small displacements in the~$q$ (position) and~$p$ (momentum) quadratures~\cite{Gottesman2001}. (We use conventions $a = (q + i p)/\sqrt 2$, $[q,p]=i$, $\hbar = 1$, vacuum variance $=1/2$.) The ideal code states of the GKP code are Dirac combs in $q$ and in $p$. Physical states are finitely squeezed approximations to these and are often modeled as a comb of Gaussian peaks of width (i.e.,~observable standard deviation)~$\sigma$, with separation~$\sqrt{\pi}$ and modulated by a larger Gaussian envelope of width~$1/\sigma$. Since these approximate states
are not orthogonal, there is a probability of misidentifying $\ket {{0}}$ as $\ket {{1}}$ (and vice versa) in a measurement of logical $Z$, which is implemented by a $q$ measurement and binning to the nearest integer multiple of~$\sqrt\pi$. Similarly, $\ket +$ and $\ket -$ may be misidentified when measuring logical $X$ with a $p$ measurement. A qubit-level measurement error occurs when the measured outcome is more than $\sqrt\pi/2$ away from the correct outcome.

{\it Gaussian quantum channel.}---%
The GKP code is tailored to combat the GQC%
, which randomly displaces the state in phase space according to a Gaussian distribution~\cite{Gottesman2001,Harrington2001}. The GQC is described by the
superoperator $\mathcal G_\xi$ acting on density operator $\rho$ as
\begin{align}
\label{eq:GQC}
\mathcal G_\xi (\rho) &= \frac{1}{\pi{\xi }^2}\int d^2\alpha\, {\rm e}^{-| \alpha |^2/{{\xi}^2}}D( \alpha ) \rho D( \alpha ) ^{\dagger },
\end{align}
where $D(\alpha) = e^{\alpha a^\dag - \alpha^* a}$ is the phase-space displacement operator. With $\alpha = \frac {1} {\sqrt 2}(u + i v)$, the position~$q$ and momentum~$p$ are displaced independently as $ q \to q + u$, $ p \to p + v $,
where $u$ and $v$ are real Gaussian random variables with mean zero and variance $\xi ^2$. Therefore, the GQC maintains the locations of the Gaussian peaks in the probability
for the measurement outcome, but it increases the variance of each spike by~$\xi ^2$ in both quadratures.

{\it Noise model.}---%
GKP error correction, in both its original~\cite{Gottesman2001} (Steane-style~\cite{Steane1997}) form and in its teleportation-based~\cite{Walshe2020} (Knill-style~\cite{Knill2005a}) form, involve measuring the deviation of the state's support in each quadrature ($q, p$) away from an integer multiple of $\sqrt\pi$. These measurement outcomes---each of the form~$s_{\rm m}= n \sqrt \pi + \Delta_{\rm m}$ with integer $n$ and $|\Delta_{\rm m}| \leq \sqrt \pi / 2$, where even and odd $n$ correspond to 0 and 1 logical bit values, respectively---together form the syndrome. Normally, each value of $\Delta_{\rm m}$ locally determines the displacement to apply in order to correct the error---either snapping back to grid in the original method~\cite{Gottesman2001} or applying a logical Pauli in the teleportation-based method~\cite{Walshe2020}. Analog QEC~\cite{Fukui2017,Fukui2018} instead feeds all these real-valued syndrome data~$s_{\rm m}$ directly to a higher-level decoder, which makes a global decision. Our proposal keeps aspects of both approaches. We use~$\Delta_{\rm m}$ to locally decide whether we keep it or give up and report the qubit as lost to the next-level decoder.

\begin{figure}[t]
\centering \includegraphics[angle=0, width=\columnwidth]{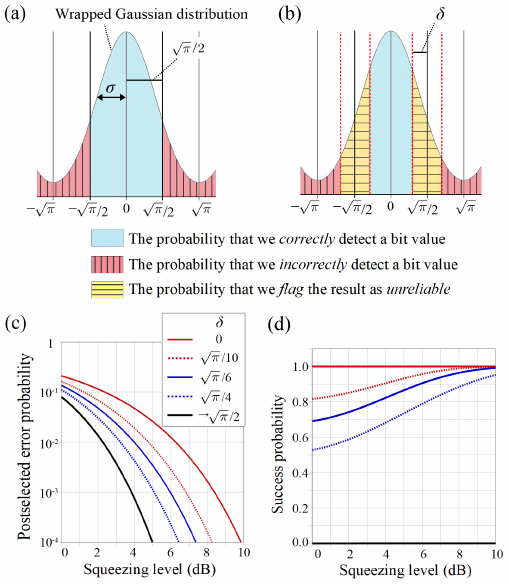} 
\caption{Effect of additive Gaussian noise~\cite{Glancy2006} on measuring a GKP qubit. (a)~%
Effect of shift by~$u \pmod {2\sqrt\pi}$, distributed according to~$p(u)$ in Eq.~\eqref{eq:pwrapped}, on an ordinary measurement of a GKP qubit~\cite{Gottesman2001}.
 (b)~%
The highly reliable measurement~\cite{Fukui2018} flags outcomes in the $2\delta$-wide ``danger zone'' (yellow) as unreliable.
(c)~Postselected error probability of the HRM
for
several values of~$\delta$.
(d)~Corresponding success probability.
Note: $(\text{Squeezing level in dB)} = -10 \log_{10}(\sigma^2/ \sigma_{\text{vac}}^2)$, where the vacuum variance~${\sigma_{\text{vac}}^2 = \tfrac 1 2}$.
}
\label{fig1}
\end{figure}

We model a damaged GKP code word as an ideal one~\cite{Gottesman2001} that has been displaced by a definite (but unknown) amount in each quadrature. This approximately models the errors due to both coherent and incoherent noise~\cite{Glancy2006,Menicucci2014,Fukui2017,Fukui2018} and simplifies the analysis. Because of the $2\sqrt\pi$-periodicity of all GKP code words, given any initial distribution $p_0(u)$ of the unknown displacement~$u$ in a single quadrature, its effect on a GKP code word is captured by folding~$p_0(u)$ into the \textit{wrapped} distribution $p(u) = \sum_{k \in \mathbb Z} p_0(u + 2k \sqrt \pi)$, whose domain is $[-\sqrt \pi , \sqrt \pi)$. When $p_0$ is a zero-mean Gaussian of variance~$\sigma^2$, it wraps into 
\begin{align}
\label{eq:pwrapped}
p(u) = \frac {1} {2\sqrt \pi}\vartheta \left( -\frac{u}{2 \sqrt{\pi }},\frac{i \sigma ^2}{2} \right)
,
\end{align}
where ${\vartheta(z,\tau) = \sum_{m \in \mathbb Z} \exp\bigl[ 2\pi i \bigl(\tfrac 1 2 m^2 \tau + m z\bigr) \bigr]}$  is a Jacobi theta function of the third kind. Figure~\ref{fig1}(a) shows this distribution and the logical effect of a shift by~${u \pmod {2 \sqrt \pi}}$ on measuring a GKP code word.

\begin{figure}[t]
\centering \includegraphics[angle=0, width=\columnwidth]{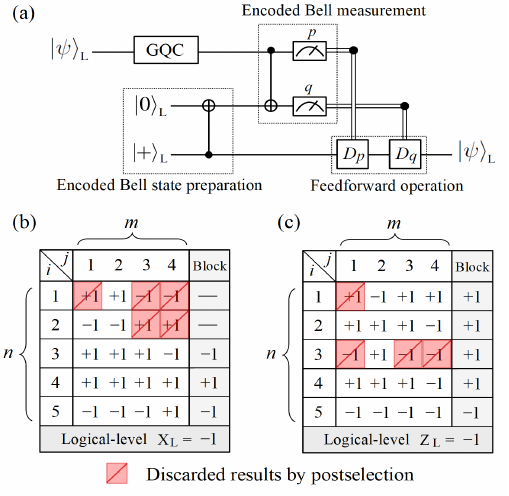} 
\caption{Trading errors for loss to improve error correction. (a) Teleportation-based QEC using GKP qubits~\cite{Gottesman2001} concatenated with the QPC~\cite{Ralph2005,Muralidharan2014}. All gates are shown at the GKP logical level. The feed forward operations~$D_{p}$ and~$D_{q}$ implement logical-level $X$ and $Z$ operations via physical-level displacements~\cite{Gottesman2001}, determined by the outcomes of the logical Bell measurement~\cite{Muralidharan2014}.
(b)~
Encoded measurement with the 
$(5,4)$ QPC
in the $X$ basis. Performing the HRM in~$p$ [Fig.~\ref{fig1}(b)] gives, for each qubit, either a binary outcome ($\pm 1$) or a located erasure error~($E$). The latter is indicated in red (with the unreliable binary outcome underneath) and occurs when the CV-level outcome is in the ``danger zone'' (see Fig.~\ref{fig1}).
A horizontal block is ignored if it has any discarded outcomes. All remaining blocks have their parity (product) taken, after which a majority vote of those parities determines the logical outcome---with heralded failure if there is no majority.
(c)~
Encoded measurement with the $(5,4)$ QPC in the $Z$ basis. This is similar to~(b), but with the HRM done in~$q$ and majority voting within a block preceding taking the parity of the blocks' voting outcomes---with heralded failure
if any block has no majority~\cite{Muralidharan2014}. (See the Supplemental Material for further details.)
Without the HRM, both logical outcomes would have been incorrect (likely due to uncorrected errors in the discarded values).%
}
\label{fig2}
\end{figure}

{\it Highly reliable measurement.}---%
Logical errors occur when the GKP syndrome value~$s_{\rm m}$, which is wrapped$\mod \sqrt \pi$, misidentifies $u$ as $u \pm \sqrt \pi$~\cite{Gottesman2001}.
The highly reliable measurement (HRM)~\cite{Fukui2018} buffers against this possibility by introducing a danger zone of outcomes~$0 \le \sqrt \pi/2 - |\Delta_{\rm m}| < \delta$ for some $\delta > 0$. Outcomes in this zone are flagged as unreliable, with $\delta \to 0$ recovering the usual case~\cite{Gottesman2001}. This corresponds to flagging as unreliable any displacement~$u \pmod{2\sqrt\pi}$ that falls within~$\delta$ of a crossover point $\pm \sqrt \pi/2$, as shown in Fig.~\ref{fig1}(b). When the HRM flags a result~$s_{\rm m}$ as unreliable, the corresponding qubit is discarded and treated as a located erasure error ($s_{\rm m} \to E$), while otherwise the result is kept and binned as usual ($s_{\rm m} \to \pm 1$)~\cite{Gottesman2001} depending on which of an even or odd multiple of $\sqrt{\pi}$ $s_{\rm m}$ is close to. The HRM is thus a ternary (three-outcome) decoder for GKP qubits that maps each raw CV outcome~$s_{\rm m}$ from $\mathbb R \to \{\pm 1,E\}$.

Given a definite displacement~$u \in [-\sqrt \pi, \sqrt \pi)$, we define probabilities for three cases: the measurement result is correct, $P^{(\correct)} = {\Pr(|u| < \sqrt \pi / 2 - \delta)}$; the result is incorrect, $P^{(\incorrect)} = {\Pr(|u| > \sqrt \pi / 2 + \delta)}$; or the result is unreliable and the qubit discarded, $P^{(\discard)} = {\Pr(- \delta < |u| - \sqrt \pi / 2 < \delta)}$. 
We further define the ``{success probability}'', $1 - P^{(\discard)}$, as the probability the qubit was not discarded and the ``{postselected error probability}'', $P_{\rm post}^{(\incorrect)} = P^{(\incorrect)}/(1 - P^{(\discard)})$, as the probability of getting an incorrect outcome within the sample of qubits that are not discarded. Decreasing the postselected error probability (by increasing~$\delta$) reduces the success probability~\cite{Fukui2018}, as shown in Figs.~\ref{fig1}(c) and~\ref{fig1}(d).

{\it Trading errors for loss}---%
The HRM is the key to improving the performance of the code against the GQC without the significant computational overhead required for conventional analog QEC for this code. Analog QEC for this code requires modeling the joint likelihood of real-valued outcomes over multimode code words, which will be intractable when the code size gets larger while the HRM maps \textit{locally detected} unreliable results to lost qubits at known locations.

Loss-tolerant QEC codes were originally proposed to overcome loss of individual photons---the main hurdle in quantum computation based on a single-photon qubit. Here, concatenating GKP qubits with one of these codes compensates for the discarded (``lost'') qubits due to using the HRM. This trade of unlocated errors for located erasures makes the logical qubit more robust.
In the following, we concatenate GKP qubits with the QPC proposed by Ralph {\it et al.}~\cite{Ralph2005} and implement
teleportation-based QEC as proposed by Muralidharan {\it et al.}~\cite{Muralidharan2014}.

The $(n,m)$ QPC~\cite{Muralidharan2014} is an $nm$-qubit code built from $n$ blocks of $m$ qubits.
Logical basis states are $
	\ket \pm_{\rm L}
=
	{
	2^{-n/2}
	\bigl(
	\ket 0^{\otimes m}
	\pm
	\ket 1^{\otimes m}
	\bigr)
	^{\otimes n}
	}
$.
In our code, the physical qubit states are square-lattice GKP states~\cite{Gottesman2001} of a single bosonic mode---i.e.,~$\ket 0 = \ket {0_{\text{GKP}}}$ and $\ket 1 = \ket {1_{\text{GKP}}}$.

We analyze the performance of our code by simulating the process shown in Fig.~\ref{fig2}, which implements error correction using the teleportation-based protocol of Ref.~\cite{Muralidharan2014} (see figure caption and the Appendix for details.) Using teleportation guarantees that the output state is already in the logical subspace, and only logical corrections are required~\cite{Knill2005a}.

{\it Numerical simulation.}---%
We evaluate our proposed QEC method using a Monte Carlo simulation of the circuit in Fig.~\ref{fig2}(a).
The input state passes through a GQC [Eq.~\eqref{eq:GQC}]. Our model simulates code-capacity noise (i.e.,~assuming no errors aside from the channel noise itself~\cite{Chamberland2017,Tuckett2020}) in order to evaluate the best possible performance of our code under the GQC and to compare with previous results~\cite{Gottesman2001,Harrington2001,Fukui2017,Fukui2018}. We consider two cases: (a)~the conventional case, which corresponds to choosing $\delta_X = \delta_Z = 0$, and (b)~an optimized choice of $\delta_X$ and $\delta_Z$ for logical $X$ and $Z$ errors (specified below).

We define the ``{failure probability}'' of the QPC as the probability that the final value of $X_{\rm L}$ or $Z_{\rm L}$ is wrong. When a heralded failure occurs---see caption of Figs.~\ref{fig2}(b) and~\ref{fig2}(c), we randomly assign the outcome. (More sophisticated handling may be possible, e.g.,~concatenating with a higher-level code.)

\begin{figure}[t]
\centering \includegraphics[angle=0, scale=1.0]{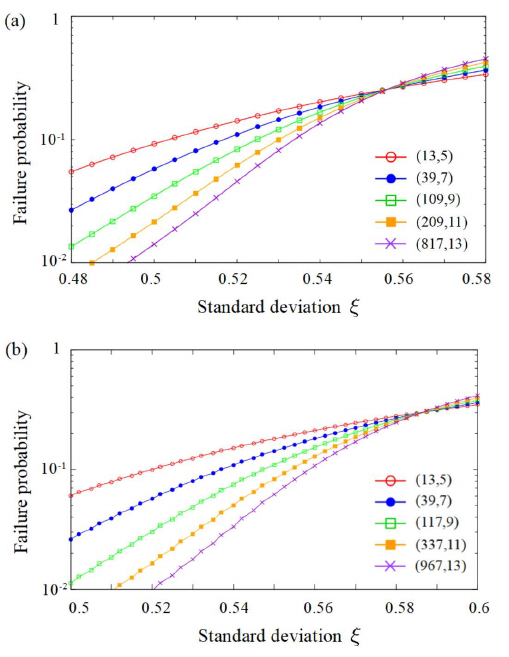} 
     \caption{Failure probabilities using the $(n,m)$ QPC, as shown in Fig.~\ref{fig2}, for (a)~$\delta_X = \delta_Z = 0$ (conventional GKP error correction~\cite{Gottesman2001}) and (b)~optimized values of the HRM parameters~$\delta_X$ and $\delta_Z$~shown in the main text.
}
\label{fig3}
\end{figure}

Figure~\ref{fig3} shows the performance of (a)~the QPC without HRM and (b)~that with HRM, as a function of the standard deviation of the GKP qubit for several sizes $(n,m)$ of the QPC. 
In Fig.~\ref{fig3}(a), we optimized the value of $n$ for the given~$m$ so that the failure probability is minimized; in Fig.~\ref{fig3}(b), we optimized $\delta_Z$ and $\delta_X$, as well as $n$, so that the failure probability is minimized.
For $m=5, 7, 9, 11, 13$, respectively, the optimized values of $n$ are $n=13, 39, 109, 209, 817$ for the conventional method~(a) and $n=13, 39, 117, 337, 967$ for our new method~(b). 
Furthermore, the optimized values of the HRM parameters $\delta_X$ and~$\delta_Z$ for~(b) are  $\delta_X/\sqrt{\pi}=$ 0.0963, 0.0967, 0.0968, 0.0968 and 0.0968,  and $\delta_Z/\sqrt{\pi}=$ 0.130, 0.134, 0.137, 0.138, and 0.139, respectively, 
which leads to the loss probability of about 15\% for $X$ and 21\%--22\% for $Z$.
The conventional method~(a) gives a threshold of $\xi \approx 0.555$, matching previous work with concatenated codes and simple decoding~\cite{Gottesman2001,Harrington2001}.
Our improved method greatly surpasses this, achieving a threshold $\xi \approx 0.585$ with a simple and efficient decoder.

{\it Conclusion.}---%
The key insight of this Letter is that one does not need to model the full likelihood function~\cite{Fukui2017,Fukui2018} to correctly interpret GKP syndrome information~\cite{Gottesman2001} within a concatenated code. Instead, the real-valued outcomes can be coarse grained to one of three qubit-level outcomes through the HRM mapping $\mathbb R \to \{\pm 1, E\}$, where $E$ represents an untrustworthy value. These ternary outcomes suffice to achieve a considerable improvement of the code against GQC by treating $E$ outcomes as erasure errors and concatenating with a qubit-level code designed to handle such errors~\cite{Ralph2005,Muralidharan2014,Munro2012,Ewert2016}.

The innovation of this work over using the full analog information for error correction~\cite{Fukui2017,Fukui2018}
lies in the efficiency and versatility of the decoder. Respectively, (1)~decoding happens in linear time since the CV-level decoding is entirely local; and (2)~the HRM wraps each GKP qubit in a simple ``{error-detecting}'' code, so concatenating with any qubit-level code designed to handle erasures~\cite{Varnava2006,Barrett2010,Tillich2013} should benefit from this type of outcome mapping. 
Further applications and extensions include improved decoding in GKP-based architectures (e.g.,~\cite{Bourassa2020,tzitrin2021fault,Larsen2021}) and in codes that exploit biased noise (e.g.,~\cite{Tuckett2020,Ataides2020,Hanggli2020,stafford2022biased}).
The numerical result implies that there is room for further improvement to achieve the hashing bound of the GQC, $\xi \approx 0.607$, by using more complicated codes, additional layers of concatenation, or sophisticated decoders.
Our innovation, however, is rooted in the simple and efficient decoder.

{\it Acknowledgments.}---We thank Joe Fitzsimons, Tim Ralph, Ben Baragiola, and Giacomo Pantaleoni for discussions. We acknowledge the organizers of the BBQ 2019 workshop, where early results of this work were presented. This work is supported by the Australian Research Council~(ARC) Centre of Excellence for Quantum Computation and Communication Technology (Project No.\ CE170100012). K.F.\ acknowledges financial support from donations from Nichia Corporation.
T.M.\ acknowledges JSPS Overseas Research Fellowships.
This work was partly supported by JST [Moonshot R$\&$D][Grant No. JPMJMS2064] and [Moonshot R$\&$D][Grant No. JPMJMS2061].

\onecolumngrid
~\\
~\\
\section*{Supplemental material for \textquotedblleft An efficient, concatenated, bosonic code for additive Gaussian noise\textquotedblright}

This supplementary material provides the details of the quantum error correction for the parity code with our method.
\\
~\\
\twocolumngrid

\subsection*{A. Bell measurement at the physical level}
\begin{figure}[t]
    \includegraphics[angle=0, width=1.0\columnwidth]{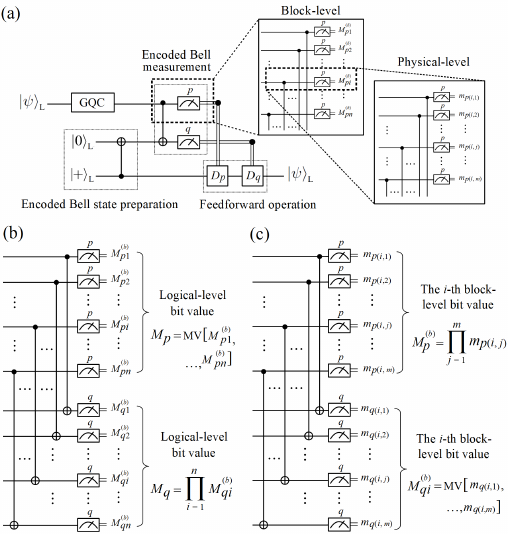}
    \caption{\label{sfig1}
A quantum circuit of the teleportation-based QEC with GKP qubits. 
(a)~The encoded data qubit~$\ket{\widetilde{\psi}}_{\rm L}$, two encoded qubits~$\ket{\widetilde{+}}_{\rm L}$, and $\ket{\widetilde{0}}_{\rm L}$ are encoded by the QPC. All gates are shown at the logical level. 
The GQC denotes the Gaussian quantum channel.
The feedforward operations~$D_{p}$ and~$D_{q}$ implement logical-level $X$ and $Z$ operations via physical-level displacements~\cite{Gottesman2001}. 
(b)~In the block-level measurement, the CNOT gate is performed on the $i^{\text{th}}$ physical qubits for the encoded data qubit and one of the encoded Bell pair. $\operatorname{MV}[ M_{p{\rm 1}}^{(b)},\cdots,M_{pn}^{(b)} ]$ represents the majority voting between the $n$ block-level qubits.
(c)~In the physical-level measurement for the $i^{\text{th}}$ block level, the CNOT gate is performed on the $j^{\text{th}}$ physical qubits in the $i$-block-level qubit for the encoded data qubit and one of the encoded Bell pair.
}   
\end{figure}

In this section, we describe the way to obtain measurement outcomes at the physical level in the encoded Bell measurement. Quantum error correction (QEC) with the quantum parity code (QPC)~\cite{Ralph2005,Muralidharan2014} is implemented by a quantum teleportation-based QEC, where the logical qubit~$\ket{\widetilde{\psi}}_{{\rm L}}$ encoded by the QPC, is teleported through the fresh encoded Bell state, as shown in Fig.~\ref{sfig1}(a). At the block level, logic-level bit values in the $X$ and $Z$ bases are obtained from the measurement outcomes of $m$ block-level bit values, respectively, as shown in Fig.~\ref{sfig1}(b). At the physical level, the $i^{\text{th}}$ block-level bit values in the $X$ and $Z$ bases are obtained from the measurement outcomes of $n$ physical-level bit values, as shown in Fig.~\ref{sfig1}(c), where the deviations of the physical GKP qubit are projected onto the ancilla physical qubit via the CNOT gate.

We here describe the physical-level Bell measurement between the physical qubits $L$ and $A$ that compose the encoded qubit and one of the encoded Bell state, respectively. At the physical level, the CNOT gate, which corresponds to the operator $\exp(-i{q}_{{\rm L}}{p}_{{\rm A}}$), transforms the quadrature operators in the Heisenberg picture as
\begin{align}
{q}_{{\rm L}} &\to   {q}_{{\rm L}},
&
{p}_{{\rm L}} &\to {p}_{{\rm L}} - {p}_{{\rm A}},  \label{eqcx} \tag{A1}\\
{q}_{{\rm A}} &\to   {q}_{{\rm A}} +  {q}_{{\rm L}},
&
{p}_{{\rm A}} &\to {p}_{{\rm A}}, \label{eqcx2}\tag{A2}
\end{align}
where ${q}_{{\rm L}}$ (${q}_{{\rm A}}$) and ${p}_{{\rm L}}$ (${p}_{{\rm A}}$) are the $q$ and $p$ quadrature operators of the physical qubit for the encoded data qubit and one of the encoded Bell state, respectively. 

Then, the The CNOT gate transforms the deviations as
\begin{align}
 {\Delta}_{q{{\rm L}}} &\to    {\Delta}_{q{{\rm L}}},
 & {\Delta}_{p{{\rm L}}} &\to {\Delta}_{p{{\rm L}}} -{\Delta}_{p{{\rm A}}}, \label{eqdis} \tag{A3}\\
 {\Delta}_{q{{\rm A}}} &\to   {\Delta}_{q{{\rm A}}} +  {\Delta}_{q{{\rm L}}}, 
 & {\Delta}_{p{{\rm A}}} &\to {\Delta}_{p{{\rm A}}},  \label{eqdis2} \tag{A4}
\end{align}
where ${\Delta}_{q{{\rm L}}}$ (${\Delta}_{q{{\rm A}}}$) and ${\Delta}_{p{{\rm L}}}$ (${\Delta}_{p{{\rm A}}}$) are the amount of the deviations of the physical qubit for the encode data qubit and one of the encoded Bell state, respectively. Since the only encoded data qubit suffer from the Gaussian quantum channel (GQC), which displaces the $q$ and $p$ quadratures randomly and independently, we assume ${{\Delta}_{q{{\rm A}}} = {\Delta}_{p{{\rm A}}} = 0}$. Thus, we obtain the outcomes of bit values $m_{p}$ and $m_{q}$ for the physical qubit, and we also obtain deviation values, $\Delta_{p{\rm m}}$ and $\Delta_{q{\rm m}}$, where ${\Delta_{p{\rm m}}={\Delta}_{p{{\rm L}}} \mod \sqrt{\pi}}$, and ${\Delta_{q{\rm m}}={\Delta}_{q{{\rm L}}} \mod \sqrt{\pi}}$ for the physical GKP qubit. We note that the bit values $m_{p}$ and $m_{q}$ are correct when the amount of deviations in $p$ and $q$ quadratures, $|{\Delta}_{p{{\rm L}}}|$ and $|{\Delta}_{q{{\rm L}}}|$, are smaller than $\sqrt{\pi}/2$, respectively.

\subsection*{B. Bell measurement at the logical level with the highly-reliable measurement}
In the decoding of the Bell measurement on the encoded qubits, we decide the encoded bit values, $M_{p}$ and $M_{q}$, by using a majority-voting procedure between the measurement outcomes~\cite{Ralph2005,Muralidharan2014}. In the following, we describe a procedure to determine $M_{p}$ and $M_{q}$ for the $(n,m)$-QPC with the highly-reliable measurement (HRM), where the HRM maps \textit{locally detected} unreliable results to lost qubits at known locations.

We describe the encoded measurement at the block level. For the encoded measurement of the $(n,m)$-QPC in the $X$ basis, the logical bit value $M_{p}$ is determined by a majority vote between $n'$ $(\leq n)$ block-level bit values, as shown in Fig.~\ref{sfig1}(b). Each block-level bit value is obtained from the product of $m$ physical-level bit values, as shown in Fig.~\ref{sfig1}(c). We here consider the $i^{\text{th}}$ $(i=1,\dotsc, n)$ block value in the $X$ basis, $M^{\rm (b)}_{p i}$, and the $j^{\text{th}}$ physical-level bit value, $m_{p(i,j)}$ $(j=1,\dotsc,m)$. Then, we define the bit values $m_{p(i,j)}=+1$ and $-1$ corresponding to $+$ and $-$ outcomes, respectively, when the HRM succeeds with probability $1 - P^{(\rm d)}$, as in the main text. When the HRM fails, with probability $P^{(\rm d)}$, the bit value is defined as $m_{p(i,j)}=0$. The block-level bit value $M^{\rm (b)}_{p i}$ is given by
\begin{equation}
M^{\rm (b)}_{p i}=\prod_{j=1}^{m} m_{p(i,j)}. \tag{B1} 
\label{eq:block_X}
\end{equation}
Therefore, when at least one of the outcome in the block is discarded, i.e., $m_{p(i,j)}=0$ for some $j$, then we discard the block-level bit value, i.e., $M^{\rm (b)}_{p i}=0$.
The probability for $M^{\rm (b)}_{p i}=0$, denoted as $P^{(\mathrm{d})}_{\rm blk X}$, is thus given by
\begin{equation}
P^{(\mathrm{d})}_{\rm blk X}=1-(1-P^{(\mathrm{d})} )^m. \tag{B2}
\end{equation}
On the other hand, if $m_{p(i,j)}\neq 0$ for any $j$ and an even number of the incorrect bit values is in the block, then we obtain the correct value for $M^{\rm (b)}_{p i}$ given in Eq.~\eqref{eq:block_X}. The probability $P^{(\mathrm{c})}_{\rm blk X}$ that we obtain the correct block-level bit value for the $X$ basis is thus given by
\begin{align}
P^{(\mathrm{c})}_{\rm blk X}&=\sum_{j=0}^{\lfloor\frac{m}{2}\rfloor} {m \choose 2j} (P^{(\mathrm{i})})^{2j} (P^{(\mathrm{c})})^{m-2j} \nonumber\\
&=\frac{(1-P^{(\mathrm{d})} )^m + (P^{(\mathrm{c})} - P^{(\mathrm{i})})^m}{2}.\tag{B3}
\end{align}
The probability $P^{(\mathrm{i})}_{\rm blk X}$ of obtaining the incorrect block-level bit value for the $X$ basis is thus given by
\begin{equation}
    P^{(\mathrm{i})}_{\rm blk X} = 1 - P^{(\mathrm{c})}_{\rm blk X} - P^{(\mathrm{d})}_{\rm blk X}. \tag{B4}
\end{equation}

Now we consider the probability $E_{\rm X}$ of obtaining the incorrect logic-level bit value for the $X$ basis.
The logical bit value is determined by using a majority vote between $n'$ $(\leq n)$ block-level bit values, where the number of discarded block-level bits is $n-n'$.
We note that the logical bit value is randomly determined when the majority voting is inconclusive---i.e., the numbers of correct and incorrect block-level bit values are the same. Thus the probability of obtaining an incorrect bit value in that case is~1/2. 
The probability $P_{\rm logic X}^{=}$ of obtaining the inconclusive logic-level bit value for the $X$ basis is given by
\begin{align}
P_{\rm logic X}^{=} &= \sum_{j=0}^{\lfloor \frac{n}{2}\rfloor} {n \choose j}{n-j \choose j}  (P^{(\mathrm{c})}_{\rm blk X})^j (P^{(\mathrm{i})}_{\rm blk X})^j (P^{(\mathrm{d})}_{\rm blk X})^{n-2j} \nonumber \\
&= (P^{(\mathrm{d})}_{\rm blk X})^{n} {}_2 F_1 \!\left(-\frac{n-1}{2},-\frac{n}{2};1 ; \frac{4 P^{(\mathrm{i})}_{\rm blk X} P^{(\mathrm{c})}_{\rm blk X}}{( P^{(\mathrm{d})}_{\rm blk X})^2} \right),\tag{B5}
\end{align}
where the hypergeometric function ${}_2F_1(a,b;c;z)$ is defined as
\begin{equation}
    {}_2F_1(a,b;c;z) = \sum_{n=0}^{\infty} \frac{(a)_n (b)_n}{(c)_n n!}z^n \tag{B6}
    \label{eq:hyper_geo}
\end{equation}
with $(x)_n = x(x+1)\cdots (x+n-1)$ being a rising factorial.
On the other hand, when the number of incorrect block-level bits is larger than the number of correct block-level bits, then the logic-level bit value for the $X$ basis is always incorrect.  The probability $P_{\rm logic X}^{>} $ that the number of incorrect block-level bits is larger than the number of correct block-level bits is given by
\begin{align}
    &P_{\rm logic X}^{>} \nonumber\\ 
     &= \sum_{j=0}^{\lfloor \frac{n}{2}\rfloor} \sum_{k=j+1}^{n - j}{n \choose j}{n-j \choose k}(P^{(\mathrm{c})}_{\rm blk X})^j (P^{(\mathrm{i})}_{\rm blk X})^k (P^{(\mathrm{d})}_{\rm blk X})^{n-j - k}  \nonumber \\
    &= \sum_{j=0}^{\lfloor \frac{n}{2}\rfloor}{n\choose j} (P^{(\mathrm{c})}_{\rm blk X})^j (1-P^{(\mathrm{c})}_{\rm blk X})^{n-j} I_{Q_{\rm blk X}}(j+1, n-2j)
        ,
    \tag{B7}
\end{align}
where $Q_{\rm blk X}\coloneqq P^{(\mathrm{i})}_{\rm blk X}/ (1-P^{(\mathrm{c})}_{\rm blk X})$, and $I_{x}(a, b)$ denotes the regularized incomplete beta function given by
\begin{equation}
    I_{x}(a, b) \coloneqq  \frac{B(x;a, b)}{B(1; a, b)}  \tag{B8}
    \label{eq:def_I}
\end{equation}
with the incomplete beta function $B(x;a, b)$ defined as
\begin{equation}
    B(x; a, b) \coloneqq \int_0^{x} t^{a-1}(1-t)^{b-1}dt. \tag{B9}
\end{equation}
Note that we used the fact that the cumulative distribution function of the binomial distribution $B(n,p)$, with its probability mass function being ${n \choose k}p^k(1-p)^{n-k}$, is given by $1 - I_{p}(k+1, n-k)$.
Thus, the logic-level error probability for the $X$ basis, $E_{\rm X}$, is given by
\begin{equation}
E_{\rm X}=\frac{1}{2} P_{\rm logic X}^{=}+P_{\rm logic X}^{>}.\tag{B10}
\end{equation}

For the encoded measurement of the $(n,m)$-QPC in the $Z$ basis, the logical bit value $M_{q}$ is determined by the product of $n$ block-level bit values as
\begin{equation}
M_{q}=\prod_{i=1}^{n} M^{\rm (b)}_{q i}, \tag{B11} \label{eq:logic_Z}
\end{equation}
and each block-level bit value, $M^{\rm (b)}_{q i}$, is obtained by a majority vote between $m$ physical-level bit values~\cite{Ralph2005,Muralidharan2014}. We here consider the $i^{\text{th}}$ $(i=1,\dotsc, n)$ block value in the $Z$ basis, $M^{\rm (b)}_{q i}$, and the $j^{\text{th}}$ physical-level bit value, $m_{q(i,j)}$ $(j=1,\dotsc,m)$. At the block level, a majority vote between $m'$ $(\leq m)$ physical-level bit values is performed, where the number of discarded block-level results is $m-m'$. 
When the majority vote is inconclusive, the block-level bit value is randomly chosen. Thus, the probability of obtaining the incorrect bit value in this case is~1/2. The probability $P^{=}_{\rm blkZ}$ of obtaining the inconclusive block-level bit value for the $Z$ basis is given by
\begin{align}
P^{=}_{\rm blkZ} &= \sum_{j=0}^{\lfloor \frac{m}{2} \rfloor} {m \choose j}{m-j \choose j} (P^{(\mathrm{c})})^{j} (P^{(\mathrm{i})})^{j} (P^{(\mathrm{d})})^{m-2j}\nonumber \\
&=(P^{(\mathrm{d})})^m {}_2 F_1 \!\left(-\frac{m-1}{2},-\frac{m}{2};1 ; \frac{4 P^{(\mathrm{i})}P^{(\mathrm{c})}}{( P^{(\mathrm{d})})^2} \right), \tag{B12}
\end{align}
where ${}_2 F_1(a,b;c;z)$ is defined in Eq.~\eqref{eq:hyper_geo}.
On the other hand, when the number of incorrect physical-level bit values is larger than that of correct bit values, then the block-level bit value for the $Z$ basis is always incorrect. The probability $P^{>}_{\rm blk Z}$ that the number of incorrect physical-level bit values is larger than that of correct ones is given by
\begin{align}
&P^{>}_{\rm blkZ} \nonumber\\
&= \sum_{j=0}^{\lfloor \frac{m}{2} \rfloor} \sum_{k=j+1}^{m-j} {m \choose j}{m-j\choose k} (P^{(\mathrm{c})})^{j} (P^{(\mathrm{i})})^{k} (P^{(\mathrm{d})})^{m-j-k} \nonumber \\
&=\sum_{j=0}^{\lfloor \frac{m}{2} \rfloor} {m\choose j}(P^{(\mathrm{c})})^{j} (1 - P^{(\mathrm{c})})^{m-j} I_{Q}(j+1, m-2j), 
\tag{B13}
\end{align}
where $Q = P^{(\mathrm{i})}/(1 - P^{(\mathrm{c})})$ and $I_x(a, b)$ is defined in Eq.~\eqref{eq:def_I}.
Then, the probability $P^{(\rm i)}_{\rm blk Z}$ of obtaining the incorrect block-level bit value for the $Z$ basis is given by
\begin{equation}
P^{(\rm i)}_{\rm blk Z} = \frac{1}{2} P^{=}_{\rm blkZ} + P^{>}_{\rm blkZ}. \tag{B14}
\end{equation}

Now, we consider the probability $E_{\rm Z}$ of obtaining the incorrect logic-level bit value for the $Z$ basis.
Since $M_{q}$ is determined by the product of $n$ block-level bit values, as shown in Eq.~\eqref{eq:logic_Z}, an even number of the incorrect block-level bit values does not change the sign. Therefore, we have
\begin{align}
E_{\rm Z}&=1 - \sum_{j=0}^{\lfloor\frac{n}{2}\rfloor} {n \choose 2j} [1-P^{(\mathrm{i})}_{\rm blk Z}]^{n-2j} [P^{(\mathrm{i})}_{\rm blk Z}]^{2j} \nonumber\\
&= \frac{1 - (1 - 2P^{(\mathrm{i})}_{\rm blk Z})^n}{2}.\tag{B15}
\end{align}
In these ways, we obtain $E_{\rm X}$ and $E_{\rm Z}$.

The overall logical error probability~$p_E$ is given by $1-(1-E_{\rm X})(1-E_{\rm Z})$, which we call the \emph{failure probability} of the QEC in the main text. To obtain the threshold value, we determine the code sizes $(n,m)$ of the QPC to ensure approximately symmetric noise at the output (i.e.,~$E_{\rm X} \sim E_{\rm Z}$), and optimize the parameter $\delta$ for the postselected measurement to minimize $p_E$ with each code size. The reason why we set $E_{\rm X} \sim E_{\rm Z}$ at the threshold value is since the logical-error probability $E_{\rm X}$ and $E_{\rm Z}$ should decrease at the same time when the code size increases. Then, it is preferable that both errors substantially decrease to zero with the code size being large enough.

\bibliography{ref.bib}

\end{document}